# Dissipative time dependent density functional theory

Roumen Tsekov
DWI, RWTH, 52056 Aachen, Germany

The simplest density functional theory due to Thomas, Fermi, Dirac and Weizsäcker is employed to describe the non-equilibrium thermodynamic evolution of an electron gas. The temperature effect is introduced via the Fermi-Dirac entropy, while the irreversible dynamics is described by a non-linear diffusion equation. A dissipative Kohn-Sham equation is also proposed, which improves the Thomas-Fermi-Weizsäcker kinetic functional.

The Schrödinger equation describes the dynamics of quantum systems. Due to mathematical difficulties, however, it is analytically solved only for a few very simple systems. The Hartree-Fock method is one of the most successful approaches to many-body problems but it neglects the particle correlations since the wave function is presented as a Slater determinate. There are many attempts to improve it [6]. During the past two decades enormous progress in the modelling of many-electron systems was achieved via the Time Dependent Density Functional Theory (TD-DFT) [6, 10], which is based on the Runge-Gross theorem [12], a non-stationary extension of the Hohenberg-Kohn theorem [5]. However, the modern TD-DFT versions do not still consider that the dynamics of many-body systems is always irreversible due to mixing trajectories. How it is known from the statistical mechanics the reduction of a many-particles probability density to single-particle distributions results in entropy production and appearance of system temperature. In the present paper an attempt is made to take into account these two effects via the simplest DFT due to Thomas, Fermi, Dirac and Weizsäcker. The temperature effect is introduced by the Fermi-Dirac entropy [11], while the irreversibility is described via a diffusion equation following from the non-equilibrium thermodynamics. Finally, a dissipative Kohn-Sham equation is also proposed, accounting for the kinetic energy better than the Thomas-Fermi-Weizsäcker density functional.

At the time of inception of quantum mechanics Thomas [13] and Fermi [4] have proposed a local density theory for description of an electron gas. In this so-called Thomas-Fermi model the kinetic energy is given by the following functional of the local electron density $\rho$

$$E_{TF}[\rho] = \int dr \rho 3\hbar^2 (3\pi^2 \rho)^{2/3} / 10m \qquad (1)$$

where $m$ is the electron mass. Since the electrons are charged particles, the electrostatic interaction between the electrons is calculated via the Hartree functional

$$E_H[\rho] = \int dr \rho \int dr' \rho e^2 /2|r-r'| \qquad (2)$$

where $e$ is the electron charge. The electrons obey, however, the Pauli principle since they are fermions. The corresponding anti-symmetry of the wave function leads to additional electrostatic effect known as exchange interaction. A simple exchange density functional

$$E_D[\rho] = -\int dr \rho 3e^2 (3\pi^2 \rho)^{1/3} /4\pi \qquad (3)$$

was introduced first by Dirac [3]. Finally, Weizsäcker [16] added a term

$$E_W[\rho] = \int dr \rho \hbar^2 [\nabla \ln(\rho)]^2 /8m \qquad (4)$$

known as the Weizsäcker correction, which improves the Thomas-Fermi kinetic energy functional. It is responsible, in particular, for the molecular bonding. In the case when the electron gas is under action of an external potential $U$, originating for instance from the interaction of the electron gas with the ions in a solid, an external potential functional can be added as well

$$E_U[\rho] = \int dr \rho U \qquad (5)$$

Thus accomplished Thomas-Fermi-Dirac-Weizsäcker model describes the electron gas at zero temperature.

How it was already mentioned, the description of a many-particles system by a local density requires a temperature. In the case of a constant temperature $T$ of the system, the Helmholtz free energy of the electron gas is the thermodynamic characteristic function. It can be presented as follows

$$F = E_{TF} + E_W + E_H + E_D + E_U - TS \qquad (6)$$

where $S$ is the entropy of the electron gas. Since the electrons are fermions, the latter can be expressed via the Fermi-Dirac functional [11]

$$S_{FD}[\rho] = -k_B \int dr \rho \ln(\rho) - k_B \int dr (\bar{\rho} - \rho) \ln(\bar{\rho} - \rho) \qquad (7)$$

where $k_B$ is the Boltzmann constant and $\bar{\rho}$ is the average electron density. Accomplishing Eq. (6) by the corresponding expressions for the energies and entropy one can calculate the local chemical potential $\mu$ of the electrons

$$\mu \equiv \frac{\delta F}{\delta N} = \frac{\hbar^2}{2m}(3\pi^2\rho)^{2/3} - \frac{\hbar^2}{2m}\frac{\nabla^2\sqrt{\rho}}{\sqrt{\rho}} + \int dr'\rho\frac{e^2}{|r-r'|} - \frac{e^2}{\pi}(3\pi^2\rho)^{1/3} + U + k_B T \ln(\frac{\rho}{\bar{\rho}-\rho}) \qquad (8)$$

where $N = \int dr\rho$ is the total number of electrons. At equilibrium the chemical potential is constant and Eq. (8) provides a mean-field Fermi-Dirac distribution [11]. When the system is not in equilibrium, the gradient of the chemical potential is the driving force for diffusion of the electrons. In the frames of the linear non-equilibrium thermodynamics the flux velocity of the electrons is proportional to this driving force, i.e. $V = -\nabla\mu/b$, where $b$ is the friction coefficient of the electrons due to their interaction with the surrounding. Substituting this relation into the continuity equation $\partial_t\rho = -\nabla\cdot(\rho V)$ yields a diffusion equation for the electrons

$$\partial_t\rho = \nabla\cdot(\rho\nabla\mu/b) \qquad (9)$$

Thus, Eq. (9) completed by the expression (8) describes the irreversible evolution of the electron gas towards thermodynamic equilibrium. As is seen, the resulting diffusion equation

$$\partial_t\rho = \nabla\cdot[\rho\nabla U_{eff}/b + \nabla\cdot(\mathbb{D}\rho)] \qquad (10)$$

is a nonlinear integro-differential one since both the effective potential

$$U_{eff}[\rho] = U + \int dr'\rho e^2/|r-r'| - e^2(3\pi^2\rho)^{1/3}/\pi \qquad (11)$$

and the effective diffusion tensor

$$\mathbb{D} = [\frac{\hbar^2}{5m}(3\pi^2\rho)^{2/3}\mathbb{I} - \frac{\hbar^2}{4m}\nabla\otimes\nabla\ln(\rho) - k_B T\frac{\bar{\rho}}{\rho}\ln(1-\frac{\rho}{\bar{\rho}})\mathbb{I}]/b \qquad (12)$$

depend on the electron density. Here $\mathbb{I}$ is the unit tensor. Note that the quantum kinetic terms contribute to $\mathbb{D}$ by increasing effectively the system temperature [15].

    Let us consider now a dilute electron gas. In this case the electron density is very low and for this reason the Thomas-Fermi, Hartree and Dirac terms can be neglected in Eq. (8), while the Fermi-Dirac entropy reduces to the Boltzmann one. The Weizsäcker correction is not negligible at low density and its corresponding potential $Q \equiv -\hbar^2\nabla^2\sqrt{\rho}/2m\sqrt{\rho}$ in Eq. (8) is, in fact, the well-known Bohm quantum potential [1]. Hence, the chemical potential from Eq. (8) acquires the simplified form

$$\mu = Q + U + k_B T \ln(\rho/\bar{\rho}) \tag{13}$$

which describes practically a single electron. Introducing expression (13) in Eq. (9) yields a thermo-quantum diffusion equation [15]

$$\partial_t \rho = \nabla \cdot [\rho \nabla (Q+U)/b + D \nabla \rho] \tag{14}$$

where $D = k_B T / b$ is the Einstein diffusion constant. The solution of Eq. (14) in the case of a free diffusing electron ($U = 0$) is a Gaussian density with dispersion $\sigma^2$ given by the expression [14]

$$\sigma^2 - \lambda_T^2 \ln(1 + \sigma^2/\lambda_T^2) = 2Dt \tag{15}$$

where $\lambda_T = \hbar/2\sqrt{mk_B T}$ is the thermal de Broglie wave length. Equation (15) is a quantum generalization of the classical Einstein law $\sigma^2 = 2Dt$ for the case of a quantum Brownian particle moving in a classical environment. It could be easily understood via the effective diffusion tensor (12), which for the present case of low Gaussian electron density simplifies to $\mathbb{D} = (D_Q + D)\mathbb{I}$. The term $D_Q \equiv \hbar^2/4mb\sigma^2$ represents the quantum diffusion coefficient and originates from the Weizsäcker correction. According to this expression for $\mathbb{D}$ the momentum dispersion of the electron is a sum of the minimal Heisenberg uncertainty $\hbar^2/4\sigma^2$ and the thermal dispersion $mk_B T$. Substituting $\mathbb{D}$ in the standard relation $\partial_t \sigma^2 \mathbb{I} = 2\mathbb{D}$ yields Eq. (15) after integration on time. At zero temperature Eq. (15) reduces to $\sigma^2 = \hbar\sqrt{t/mb}$ [14].

It is well known that the Thomas-Fermi model is not very accurate and the main reason for inaccuracy is the kinetic energy functional. It is possible to improve the correctness of the Thomas-Fermi-Weizsäcker model by explicit calculation of the kinetic energy via the DFT Schrödinger equation introduced by Kohn and Sham [7]. Thus, accomplishing it by the Fermi-Dirac entropy and the electron friction one can propose a dissipative Kohn-Sham equation

$$i\hbar \partial_t \phi_k = [-\hbar^2 \nabla^2/2m + U_{eff} + k_B T \ln(\frac{\rho}{\bar{\rho}-\rho}) - i\hbar b \ln(\phi_k/\phi_k^*)/2]\phi_k \tag{16}$$

where the electron density is given by $\rho = \sum \phi_k^* \phi_k$. One can enhance Eq. (16) additionally by more sophisticated expressions [6] for the exchange potential in Eq. (11). The last frictional term in Eq. (16) is introduced first by Kostin [8] and corresponds to the Ohmic dissipation. Thus, the electron density obtained from Eq. (16) exhibits irreversible evolution to thermodynamic equilibrium. In the case of a single electron Eq. (16) reduces to [2]

$$i\hbar\partial_t\phi = [-\hbar^2\nabla^2/2m + U + k_BT\ln(\phi^*\phi) - i\hbar b\ln(\phi/\phi^*)/2]\phi \qquad (17)$$

Introducing the Madelung transformation [9] of the wave function $\phi = \sqrt{\rho}\exp(im\int dr\cdot V/\hbar)$, Eq. (17) can be split into the following two equations

$$\partial_t\rho = -\nabla\cdot(\rho V) \qquad m\partial_t V + mV\cdot\nabla V + bV = -\nabla[Q + U + k_BT\ln(\rho)] \qquad (18)$$

The first one is the continuity equation, while the second equation is the macroscopic force balance. In the strong friction limit one can neglect the first two inertial terms to obtain the linear non-equilibrium thermodynamic balance $V = -\nabla[Q + U + k_BT\ln(\rho)]/b = -\nabla\mu/b$. Substitution of this expression for $V$ in the continuity equation above results in Eq. (14). Hence, the latter is not affected by the approximations made in the Thomas-Fermi model. It should be noted, however, that Eq. (14) is still approximate since it does not provide the exact quantum canonical Gibbs distribution as equilibrium solution. After Bohm [1] the quantum potential is understood as additional potential energy of the electrons. However, how it was shown in the present paper $Q$ originates from the Weizsäcker correction and hence its origin is kinetic. Since the Weizsäcker correction is proportional to the Fisher information, the quantum potential is the chemical potential corresponding to the Fisher entropy $S_F = -\int dr\rho\nabla^2\ln(\rho)$.

## Appendix

The fact that the quantum potential is a kind of chemical potential is evident from the hydrostatic balance. Integrating $\nabla\cdot\mathbb{P} = \rho\nabla Q$ yields an expression $\mathbb{P} = -(\hbar^2/4m)\rho\nabla\otimes\nabla\ln\rho$ [9] for the quantum pressure tensor. Thus, the chemical potential of a particle in vacuum reads $\mu = Q + U$ (13) and it satisfies the Gibbs-Duhem equation $\nabla\mu = \rho^{-1}\nabla\cdot\mathbb{P} + \nabla U$ at zero temperature. Now, we are in a position to define the particle internal energy $\varepsilon$ via the first law of thermodynamics $\nabla\varepsilon = -\mathbb{P}\cdot\nabla\rho^{-1} + \nabla U$. It is possible to integrate this equation and the result

$$\varepsilon = -\hbar^2(\nabla\ln\rho)^2/8m + U \qquad (19)$$

shows that the energy of a particle is given by the Weizsäcker correction [16] but with negative sign. It is evident that the difference $\mu - \varepsilon = \text{tr}(\mathbb{P})\rho^{-1}$ between the chemical potential and the internal energy is equal to the product of the pressure and volume per a particle, as required.

According to thermodynamics, at equilibrium the chemical potential $\mu = Q + U$ is constant, which corresponds straightforward to the Schrödinger equation. Therefore, the eigenvalues of the Schrödinger equation are Gibbs energies, i.e. they are not internal energies of the system. To demonstrate the importance of this new detail, let us consider the very controver-sial case of a quantum harmonic oscillator. In the ground state the corresponding probability density, obtained from the Schrödinger equation, reads $\rho = \sqrt{m\omega_0 / \pi\hbar} \exp(-m\omega_0 x^2 / \hbar)$, where $\omega_0$ is the oscillator own frequency. Using this density one can calculate the oscillator chemical potential $\mu = Q + U = \hbar\omega_0 / 2$ and its internal energy from Eq. (19) $\varepsilon = 0$. As is seen, the oscillator internal energy is zero and the so-called zero point energy $\hbar\omega_0 / 2$ of the oscillator is, in fact, the chemical potential of the oscillator in vacuum. It represents the energy to place the oscillator in vacuum and $\mu$ could also vanish if one is able to extract the oscillator from vacuum. This shows again that the vacuum fluctuations are the reason for quantum mechanics.